\documentclass[prd,floatfix,showpacs,eqsecnum,nofootinbib,superscriptaddress,]{revtex4}
\usepackage{epsfig}
\usepackage{amsmath}
\usepackage{mathtools}

\newcommand{\eq}[1]{Eq.~(\ref{#1})}
\newcommand{\fig}[1]{Fig.~{\ref{#1}}}

\newcommand{\be}{\begin{equation}}
\newcommand{\ee}{\end{equation}}
\newcommand{\bea}{\begin{eqnarray}}
\newcommand{\eea}{\end{eqnarray}}
\newcommand{\ben}{\begin{eqnarray*}}
\newcommand{\een}{\end{eqnarray*}}

\newcommand{\DS}{Dyson-Schwinger }
\newcommand{\BS}{Bethe-Salpeter }
\newcommand{\ST}{Slavnov-Taylor }
\newcommand{\YM}{Yang-Mills }

\newcommand{\w}{\omega}
\newcommand{\e}{\varepsilon}
\newcommand{\al}{\alpha}
\newcommand{\ba}{\beta}
\newcommand{\ga}{\gamma}
\newcommand{\G}{\Gamma}
\newcommand{\de}{\delta}

\newcommand{\si}{\sigma}

\newcommand{\ov}[1]{\overline{#1}}
\newcommand{\dk}[1]
{\,\,\,\raisebox{-0.4ex}{\large $\bar{}$}\!\!d\,{#1}\,}

\begin{document}

\title{Heavy tetraquark confining potential in Coulomb gauge QCD}
\author{Carina~Popovici}
\affiliation{
Institut f\"ur Theoretische Physik, Justus-Liebig-Universit\"at Giessen,
 35392 Giessen, Germany }
\affiliation{Institut f\"ur Physik, Karl-Franzens-Universit\"at Graz, A-8010
  Graz, Austria }
\author{Christian~S.~Fischer}
\affiliation{
Institut f\"ur Theoretische Physik, Justus-Liebig-Universit\"at Giessen,
 35392 Giessen, Germany }
\date{\today}
\begin{abstract}
We present an analytic nonperturbative solution of the Yakubovsky equation for
tetraquark states in the case of equal separations and energies, and
demonstrate a direct connection between the tetraquark confinement potential
and the temporal gluon propagator. To this end we employ a leading-order heavy
quark mass expansion of the Coulomb gauge QCD action, and use the dressed
two-point functions of the Yang-Mills sector only. As a result we find a
bound state energy that rises linear with distance and a string tension twice
as large as in a $q\bar{q}$-system.
\end{abstract}
\maketitle

\section{Introduction}
\setcounter{equation}{0}

Exotic states in the heavy quark sector are an increasingly fascinating
subject to study. With the discovery and confirmation of many new XYZ-states
at BaBar, Belle, LHC and BES, interpretations of some of these clearly favor
states which go beyond the time-honored classification of hadrons into mesons
and baryons \cite{Brambilla:2010cs,Bodwin:2013nua}, opening up the exciting
possibility of the identification of tetraquark, meson molecule or hybrid
states.  The idea of tetraquarks is around for quite some time. For the light
quark sector, Jaffe proposed that the light scalar nonet including $f_0(980)$
and $a_0(980)$ can be interpreted as a $qq\bar q\bar q$ state instead of
$q\bar q$ \cite{Jaffe:1976ig,Jaffe:1976ih}. Indeed, the mass ordering and
decay patterns of these states nicely fit this picture. In the heavy quark
sector, charged states like the $Z^+_c(3900)$ or the $Z^+_b(10610)$ and their
cousins cannot be identified with ordinary quarkonia and therefore strongly
suggest an interpretation in terms of tetraquarks.

Theoretically, tetraquarks can be described by a generalized \BS equation for
four particle states, originally proposed by Yakubovsky
\cite{Yakubovsky:1966ue} (see also Refs.~\cite{Gloeckle,Fonseca:1986rw} for
pedagogical introductions).  In a covariant setting, this equation, rounded
off to account for quantum-field theoretical effects
\cite{Khvedelidze:1991qb}, has been solved under the approximation that the
$4q$ state is described by a coupled system of two-body equations with meson
and diquark constituents \cite{Heupel:2012ua}. Based on previous
investigations, which showed that a rainbow-ladder kernel is most robust in
meson \BS calculations \cite{Fischer:2006ub,Fischer:2005en,Watson:2004kd},
results of tetraquark masses have been obtained by employing a
phenomenologically validated one gluon-exchange interaction. A complete
classification of tetraquark states in terms of spin-flavor, color and spatial
degrees of freedom has been constructed in \cite{Santopinto:2006my}.  Other
investigations of tetraquark states include large $N$-limit calculations
\cite{Weinberg:2013cfa}, effective theory studies
\cite{Black:1998wt,Maiani:2004uc,Giacosa:2006tf} and relativistic quark models
\cite{Ebert:2005nc}.

From a fundamental perspective, tetraquarks offer interesting insights into
the underlying structure of the strong interaction.  The relationship between
the non-perturbative scale associated with confinement (the string tension)
and the gluon sector is of crucial importance in understanding the low-energy
properties of QCD. On the lattice, Wilson loops exhibit an area law at
intermediate distances that corresponds to a linearly rising potential,
whereas the corresponding coefficient, so-called Wilsonian string tension, can
be explicitly related to a hadronic scale \cite{Sommer:1993ce}. Within
continuous functional approaches, investigations carried out in Coulomb gauge
have shown that in the heavy quark sector (and at least under truncation) one
can identify a direct connection between the temporal \YM Green's function and
the potential that confines quarks, both in the two- and three-body case
\cite{Popovici:2010mb,Popovici:2010ph}.  In the Hamiltonian formalism, the
physical string tension can be related to both the temporal Wilson loop
\cite{Pak:2009em} and non-abelian color Coulomb potential \cite{Epple:2006hv}.

While the potential that confines two and three quarks has been relatively
extensively studied with continuous methods as well as on the lattice (see for
example Ref.~\cite{Popovici:2013fya} for a review), the interaction between
quarks in a 4$q$ system has received little attention.  On the lattice, the
problem of van der Waals forces has been investigated, and it has been shown
that a flux tube recombination takes place, i.e., around a level-crossing
point, the confining potential flips between the disconnected 'two-meson'
Ansatz and the state where the quarks and antiquarks are connected by a
double-Y shaped flux tube, and this implies that the van der Waals forces are
absent at long distances \cite{Okiharu:2004ve}. Continuum studies that have
investigated the absence of long range forces in tetraquarks include
Refs.~\cite{Appelquist:1978rt, Feinberg:1979yw}.

In this work we will study the nature of the confining force in tetraquarks
using a framework gauge fixed to Coulomb gauge.  The realization of
confinement in Coulomb gauge centers around the Gribov-Zwanziger scenario,
which conjectures that the confining potential is provided by the temporal
gluon propagator, whereas the spatial propagator is suppressed at long
distances \cite{Gribov:1977wm}.  In addition, this gauge possesses a number of
features that recommend it as an appropriate tool to study the low-energy
sector of QCD: within the first order formalism, the total charge of the
system is conserved and vanishing, the system reduces naturally to the
physical degrees of freedom \cite{Zwanziger:1998ez}, and the problem of
divergent energy integrals disappears \cite{Reinhardt:2008pr}. In Coulomb
gauge, the \DS equations for both \YM and quark sectors have been derived, and
perturbative results have been obtained
\cite{Watson:2006yq,Watson:2007vc,Watson:2007mz,Popovici:2008ty}.  On the
lattice, one important result (which shall be extensively used in this work)
is that the temporal gluon propagator is energy-independent, and it behaves
like $1/\vec q^4$ for vanishing $\vec q$
\cite{Cucchieri:2000hv,Quandt:2008zj}. The lattice results agree with the
analytical findings obtained from the Hamiltonian approach to \YM theory
\cite{Epple:2006hv, Epple:2007ut, Szczepaniak:2003ve}.

This paper is a natural continuation of previous works including one of the
authors \cite{Popovici:2010mb,Popovici:2010ph}, where meson and baryon bound
states have been investigated via \BS and Faddeev equations, respectively.
Based on a leading-order expansion in the heavy quark mass originally
developed within heavy quark effective theory [HQET] \cite{Neubert:1993mb}, a
direct connection between the temporal gluon propagator and the string tension
has been derived.\footnote{The heavy quark limit has also been recovered under
  a (perturbative) leading order truncation of \DS equations
  \cite{Watson:2012ht,Watson:2011kv}.} Here we follow the same approach and
consider the Yakubovsky equation for four-quark states
\cite{Yakubovsky:1966ue} in Coulomb gauge at leading order in the mass
expansion, in the symmetric case (i.e., the separation between quarks are
equal), at equal energies, and by including only 2PI contributions. We will
employ lattice results for the temporal gluon propagator, and in addition, we
will use previous findings, namely that the kernel of the \BS equation reduces
non-perturbatively to the ladder truncation. In this setting, we will provide
an exact analytical solution to the Yakubovsky equation, which then naturally
leads to the confining potential of a $4q$ system.

The organization of this paper is as follows. In Sec.~II we briefly
survey the results obtained for heavy quark systems. We review the
main steps of the expansion of QCD action in powers of the inverse
quark mass, and discuss the results obtained for the heavy quark
propagator and the corresponding (temporal) quark-gluon vertex. In
Sec.~III we present the Yakubovsky equation for tetraquark
states. Similar to the case of meson and baryon states, we establish
(at least under truncation) a direct relation between the physical
string tension and the temporal component of the gluon propagator. A
short summary and conclusions are presented in Sec.~IV.

\section{Expansion in the heavy quark mass}

In this section we outline the results obtained within heavy quark limit that
are relevant for this work, and direct the reader to
Ref.~\cite{Popovici:2010mb} for a full account. We employ the standard
notations and conventions: spatial indices are labeled with roman letters $i$,
$j$,\ldots, and the superscripts $a$, $b$,\ldots denote color indices in the
adjoint representation; flavor, Dirac spinor and (fundamental) color indices
are commonly denoted with an index $\al, \ba\dots$.  We work in Minkowsky
space, with the metric $g_{\mu\nu}=\mbox{diag}(1,-\vec{1})$.  The Dirac
$\ga$-matrices satisfy $\{\ga^\mu,\ga^\nu\}=2g^{\mu\nu}$, where the notation
$\ga^{i}$ refers to the spatial component and the minus sign arising from the
metric has been explicitly taken into account. $f^{abc}$ are the structure
constants of the $SU(N)$ group, with the Hermitian generators $[T^a,T^b]=i
f^{abc}T^c$ normalized via $\mbox{Tr}(T^aT^b)=\de^{ab}/2$, and the Casimir
factor $C_F=(N^2-1)/2N $.

The idea that underlies the heavy quark mass expansion is the so-called heavy
quark decomposition, i.e., the (full) quark field $q_\al$ is separated into
two components via the spinors $h$ and $H$ as follows:
\be
q_\al(x)=e^{-i mx_0}\left[h(x)+H(x)\right]_\al,\;
h_\al(x)=e^{i mx_0} \frac{\openone+\ga^0}{2} q_\al(x),\;
H_\al(x)=e^{i mx_0}\frac{\openone-\ga^0}{2} q_\al(x)
\label{eq:qdecomp}
\ee
(similarly for the antiquark field).  In our Coulomb gauge functional
approach, this particular type of heavy quark transform adapted from HQET
\cite{Neubert:1993mb} can be simply regarded as an arbitrary decomposition.
This is then inserted into the QCD generating functional, and, after
integrating out the $H$-fields, an expansion in the heavy quark mass is
performed (throughout this work we shall use the established terminology
``mass expansion'', instead of ``expansion in the \emph{inverse} quark
mass''). We mention here that the quark and antiquark sources are kept in all
steps of the calculation, such that the full gap and Yakubovsky equations can
be employed, whereas the kernels, propagators and vertices are replaced with
their expressions at leading order in the mass expansion. The decomposition
\eq{eq:qdecomp} leads to the suppression of the spatial gluon propagator at
leading order in the mass expansion, which in turn means that at LO the
attached gluons couple to the constituent quarks of the four-body state via a
temporal quark-gluon vertex.  We refer the reader to
Ref.~\cite{Popovici:2008ty} for a detailed discussion regarding source terms
and the expansion of the QCD action in the parameter $1/m$.

Before we provide our solution for the heavy quark propagator, it is
appropriate to briefly discuss our truncation scheme, which has also been
employed in \cite{Popovici:2010mb,Popovici:2010ph}. In the context of the
heavy mass expansion, we restrict ourselves to dressed two-point functions of
the \YM sector (i.e., the nonperturbative gluon propagators) and set all the
pure \YM vertices and higher $n$-point functions occurring in the quark
equations to zero. This truncation is justified by the fact that the total
number of loops containing \YM vertices is drastically reduced: on the one
hand, these vertices only contribute at second order perturbatively, since the
leading order perturbative corrections containing purely temporal vertices
vanish (temporal \YM vertices are zero at tree-level \cite{Watson:2007vc}),
and on the other hand, loops containing spatial \YM vertices are suppressed by
the mass expansion.

From the full Coulomb gauge gap equation (i.e., first order formalism without
mass expansion \cite {Popovici:2008ty}), supplemented by the \ST identity, we
find the following solution for the heavy quark propagator,
\bea
W_{\ov{q}q\al\ba}(k_0)=\frac{-i\de_{\al\ba}}{\left[k_0-m-
{\cal I}_r+i\e\right]}+{\cal O}\left(1/m\right),
\label{eq:quarkpropnonpert}
\eea
with
\bea
{\cal I}_r =\frac{1}{2}g^2C_F
\int_r\frac{\dk{\vec{\w}}
D_{\si\si}(\vec{\w})}{\vec{\w}^2}
+{\cal O}\left(1/m\right).
\label{eq:inco}
\eea
The constant ${\cal I}_r$ is implicitly regularized, under the assumption that
the order of the integration is set such that the temporal integral is
performed first, and the spatial integral is regularized and finite. The
non-perturbative temporal gluon propagator entering ${\cal I}_r$ is given by:
\be
W_{\si\si}^{ab}(\vec k)=
\de^{ab}\frac{i}{\vec{k}^2}D_{\si\si}(\vec{k}^2).
\label{eq:Wsisi}
\ee
Following lattice results \cite{Quandt:2008zj}, and also continuum
investigations \cite{Cucchieri:2000hv}, we assume that the dressing
function $D_{\si\si}$ is energy independent and diverges like
$1/\vec{k}^2$ in the infrared. From the \ST identity, combined with
the solution \eq{eq:quarkpropnonpert}, one easily finds that the
temporal quark-gluon vertex remains non-perturbatively bare,
\be
\G_{\ov{q}q\si\al\ba}^{a}(k_1,k_2,k_3)=
\left[gT^a\right]_{\al\ba}+{\cal O}\left(1/m\right),
\label{eq:feyn}
\ee
whereas the spatial vertex is subleading in the heavy mass expansion
\cite{Popovici:2010mb}.  The heavy quark propagator
\eq{eq:quarkpropnonpert} has a few remarkable properties which we
shall discuss here briefly. Firstly, as a result of the mass
expansion, this propagator has a single pole in the complex
$k_0$-plane, as opposed to the conventional Feynman quark
propagator. Therefore, it is necessary to explicitly define the
Feynman prescription. It then follows that the closed quark loops
vanish due to energy integration,
\be
\int\frac{dk_0}{\left[k_0-m- {\cal I}_r+i\e\right]
\left[k_0+p_0-m- {\cal I}_r+i\e\right]}=0,
\label{eq:tempint}
\ee
meaning that the theory is quenched at lowest order in the heavy quark mass
expansion.  A further observation is that the propagator
\eq{eq:quarkpropnonpert} is diagonal in the outer product of the fundamental
color, flavor and spinor spaces, due to the decoupling of the spin degree of
freedom in the heavy quark limit \cite{Neubert:1993mb}. Finally, the position
of the pole in the heavy quark propagator does not have a physical meaning
since the quark cannot be on-shell. As soon as the regularization is removed,
the poles are shifted to infinity, and this implies that only the relative
energy plays a role in a hadronic system (or, if a single quark is considered,
one needs infinite energy to create it from the vacuum). Indeed, it has long
been known that the absolute energy does not have a physical meaning, and that
only the relative energy, which in the case of tetraquarks is derived from the
Yakubovsky equation, must be considered \cite{Adler:1984ri}.

Since the heavy quark mass expansion breaks the charge conjugation
symmetry, the antiquark and quark propagators are not equivalent. The
Feynman prescription for the antiquark propagator is derived from the
observation that the \BS equation must have a physical interpretation
of bound states -- in this case, the quark and the antiquark are not
connected by a primitive vertex, and hence they do not create a
virtual quark-antiquark pair (closed loop) but a system composed of
two separate unphysical particles. For the antiquark propagator we
obtain (the derivation is similar to the quark propagator):
\bea
W_{q\ov{q}\al\ba}(k_0)=\frac{-i\de_{\al\ba}}{\left[k_0+m-
{\cal I}_r+i\e\right]}+{\cal O}\left(1/m\right),
\label{eq:antiquarkpropnonpert}
\eea
and the corresponding vertex is given by:
\be
\G_{q\ov{q}\si\al\ba}^{a}(k_1,k_2,k_3)=
-\left[gT^a\right]_{\ba\al}+{\cal O}\left(1/m\right).
\label{eq:feyn1}
\ee
A last important consequence of the heavy quark mass expansion is the
reduction of the interaction kernels from both \BS and Yakubovsky equations to
ladder exchange, since the crossed box contributions cancel due to energy
integration over multiple quark propagators with the same Feynman prescription
(see Ref.~\cite{Popovici:2010mb} for a detailed discussion and calculation).

\section{Yakubovsky equation for tetraquark states}

\begin{figure}[t]
\vspace{0.5cm}
\includegraphics[width=0.85\linewidth]{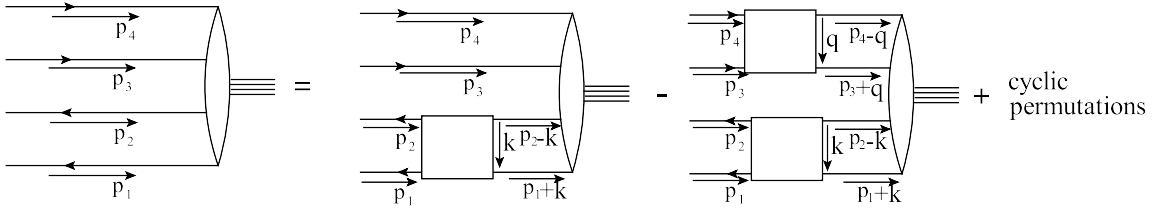}
\caption{\label{fig:tetraquark}
 Yakubovsky equation for four quark bound states. Solid lines represent the
 quark propagator, and boxes represent the meson, diquark and anti-diquark
 kernels, respectively. The ellipse depicts the Yakubovsky vertex function
 corresponding to the bound state represented by a quadruple-line.}
\end{figure}

The Yakubovsky equation is a four-body bound state equation which has
been successfully used to describe tetraquark states. It embodies
two-, three-and four-quark irreducible diagrams. Since the irreducible
three- and four-body forces all involve pure \YM vertices which, as
discussed in the previous section, are neglected in our approach, it
follows that there are no 3PI and 4PI contributions at this level of
approximation. Consequently, it is appropriate to employ only the
(anti)diquark and meson kernels that also appear in the corresponding
\BS equations. In a covariant setting, a similar approximation has
been successfully applied to study tetraquark bound states
\cite{Heupel:2012ua}.

In this approximation, and by using the formulation \cite{Khvedelidze:1991qb},
where the correct multiplicity is taken into account via the inclusion of
two-pair kernels, the Yakubovsky equation reads
\bea
\lefteqn{\G_{\al\ba\ga\de}(p_1,p_2,p_3,p_4) }\nonumber\\
&=&\int\dk{k}\left\{
S^{(qq)}_{\ba\al;\al^{\prime}\ba^{\prime}} (p_1,p_2;k)
\,\G_{\al^{\prime}\ba^{\prime}\ga\de}(p_1+k,p_2-k,p_3,p_4)
+S^{(aa)}_{\de\ga;\ga^{\prime}\de^{\prime}} (p_3,p_4;k)
\,\G_{\al\ba\ga^{\prime}\de^{\prime}}(p_1,p_2,p_3+k,p_4-k)\right.
 \nonumber\\
&+&S^{(qa)}_{\ga\ba;\ba^{\prime}\ga^{\prime}}(p_2,p_3;k)
\,\G_{\al\ba^{\prime}\ga^{\prime}\de}(p_1,p_2+k,p_3-k,p_4) 
+S^{(qa)}_{\de\al;\al^{\prime}\de^{\prime}}(p_1,p_4;k)
\,\G_{\al^{\prime}\ba\ga\de^{\prime}}(p_1-k,p_2,p_3,p_4+k)
\nonumber\\
&+&\left.  S^{(qa)}_{\ga\al;\al^{\prime}\ga^{\prime}} (p_1,p_3;k)
\,\G_{\al^{\prime}\ba\ga^{\prime}\de}(p_1+k,p_2,p_3-k,p_4)
+S^{(qa)}_{\de\ba;\ba^{\prime}\de^{\prime}} (p_2,p_4;k)
\,\G_{\al\ba^{\prime}\ga\de^{\prime}}(p_1,p_2+k,p_3,p_4-k)
\right\}\nonumber\\
&-&\int \dk{k}\dk{q}
\left\{
S^{(qq)}_{\ba\al;\al^{\prime}\ba^{\prime}} (p_1,p_2;k)\,
S^{(aa)}_{\ga\de;\de^{\prime}\ga^{\prime}} (p_3,p_4;q)
\,\G_{\al^{\prime}\ba^{\prime}\ga^{\prime}\de^{\prime}}
(p_1+k,p_2-k,p_3+q,p_4-q) \right. \nonumber\\
&+&
S^{(qa)}_{\ga\al ;\al^{\prime}\ga^{\prime}}(p_1,p_3;k)\,
S^{(qa)}_{\de\ba;\ba^{\prime}\de^{\prime}}(p_2,p_4;q)
\,\G_{\al^{\prime}\ba^{\prime}\ga^{\prime}\de^{\prime}}
(p_1+k,p_2-q,p_3-k,p_4+q)
 \nonumber\\
&+&\left. 
S^{(qa)}_{\de\al;\al^{\prime}\de^{\prime}}(p_1,p_4;k)\,
S^{(qa)}_{\ga\ba;\ba^{\prime}\ga^{\prime}}(p_2,p_3;q)
\,\G_{\al^{\prime}\ba^{\prime}\ga^{\prime}\de^{\prime}}
(p_1+k,p_2+q,p_3-q,p_4-k)
\right\}.
\label{eq:bs1}
\eea
The amplitudes $S$ contain three types of kernels, corresponding to
quark-quark, antiquark-antiquark and quark-antiquark pairs. As discussed in
the previous section, in the limit of heavy quark mass these kernels reduce to
ladder gluon exchange:
\bea
S^{(qq)}_{\al\ba;\ba^{\prime}\al^{\prime}} (p_i,p_j;k)
&=&
g^2 T_{\al\al^{\prime}}^{a}T_{\ba\ba^{\prime}}^{a} W_{\si\si}(\vec k)
W_{\bar qq}(p_i^0+k_0)W_{\bar qq}(p_j^0-k_0)\nonumber\\
S^{(aa)}_{\al\ba;\ba^{\prime}\al^{\prime}} (p_i,p_j;k)
&=&g^2 T_{\al^{\prime}\al}^{a}T_{\ba^{\prime}\ba}^{a} W_{\si\si}(\vec k)
W_{q\bar q}^T(p_i^0+k_0)W_{q\bar q}^T(p_j^0-k_0)
\nonumber\\
S^{(qa)}_{\al \ba;\ba^{\prime}\al^{\prime}} (p_i,p_j;k)
&=&
-g^2 T_{\al\al^{\prime}}^{a}T_{\ba^{\prime}\ba}^{a} W_{\si\si}(\vec k) 
W_{q\bar q}^T(p_i^0+k_0)W_{\bar qq}(p_j^0-k_0).
\label{eq:bspairs}
\eea
In the above, we have already replaced the temporal quark-gluon
vertices by their expressions Eqs.~(\ref{eq:feyn}) and
(\ref{eq:feyn1}).  In our convention, $p_1,p_2$ denote the quark
momenta, $p_3, p_4$ the antiquark momenta, and $P_0=
{\scriptstyle\sum\limits_{i=1}^4}p_i^0$ is the pole four-momentum
(total energy) of the bound tetraquark state.  $\G_{\al\ba\ga\de}$
represents the quark-tetraquark vertex for a particular bound state
and its indices denote explicitly only its quark content. Just like
the heavy quark propagator, $\G_{\al\ba\ga\de}$ becomes a Dirac scalar
due to the decoupling of the spin in the heavy mass limit.  Similar to
the homogeneous \BS equation for mesons and Faddeev equation for
baryons, the integral equation (\ref{eq:bs1}) depends only
parametrically on the total energy $P_0$ (for notational convenience
we have dropped the $P_0$ dependence from $\G_{\al\ba\ga\de}$). We
also note that, as in the case of meson and baryon bound states, the
energy independence of the temporal gluon propagator will play a key
role in the derivation of the confining potential. The Yakubovsky
equation is diagrammatically shown in \fig{fig:tetraquark}.

Before we specify our Ansatz for the Yakubovsky vertex, let us shortly recall
the energy behavior of the meson and baryon vertices. Whereas in the case of
mesons it was straightforward to show that the \BS vertex was
energy-independent, the quark-baryon vertex did contain an energy component,
similar in structure to the quark propagator \cite{Popovici:2010ph}. In case
of tetraquarks the relative energy does not cancel, hence it is reasonable to
assume that the Yakubovsky vertex obeys a separable \emph{Ansatz} (recall that
the dependence on the total energy is implicit):
\be
\G_{\al\ba\ga\de}( p_1, p_2,  p_3,p_4)=\Psi_{\al\ba\ga\de}
\G_{t}( p_1^0, p_2^0, p_3^0,p_4^0)
\G_{s}( \vec p_1,\vec  p_2, \vec  p_3,\vec p_4).
\label{eq:qqqq}
\ee
$\G_{t}$ and $\G_{s}$ represent the temporal and spatial component,
respectively, and $\Psi_{\al\ba\ga\de}$ denotes the color component, with
$\al,\ba$ being quark, and $\ga, \de$ antiquark indices.  

The color structure of tetraquarks is nontrivial since a singlet can be
obtained via two different representations \cite{Santopinto:2006my}.
Interestingly, diquark and antidiquark pairs also contribute to the color
singlet tetraquark state, although themselves cannot exist as color singlets
(at least for $N=3$ colors). By writing the color function $\Psi$ as
\be
\Psi_{\al\ba\ga\de}=
\de_{\al\de}\de_{\ba\ga}+\de_{\al\ga}\de_{\ba\de},
\ee
and with the Fierz identity for the generators
\be
2\left[T^a\right]_{\al\ba}\left[T^a\right]_{\de\ga}=\de_{\al\ga}
\de_{\de\ba}-\frac{1}{N}\de_{\al\ba}\de_{\de\ga},
\label{eq:fierz}
\ee 
we calculate the color factors corresponding to various channels:
\bea
T^a_{\al\al^{\prime}}T^a_{\ba\ba^{\prime}}
\Psi_{\al^{\prime}\ba^{\prime}\ga\de}&=&
\frac{1}{2} \left(1-\frac{1}{N}\right) 
\Psi_{\al\ba\ga\de}
\textrm{~~(diquark and antidiquark
  channel)}\nonumber\\
T^a_{\al\al^{\prime}}T^a_{\ga^{\prime}\ga}
\Psi_{\al^{\prime}\ba\ga^{\prime}\de}&=&
\frac{1}{2} \left(1+N-\frac{2}{N}\right)
\Psi_{\al\ba\ga\de}\textrm{~~(meson  channel)}
\nonumber\\
T^a_{\al\al^{\prime}}T^a_{\ba\ba^{\prime}}
T^b_{\de^{\prime}\de}T^b_{\ga^{\prime}\ga}
\Psi_{\al^{\prime}\ba^{\prime}\ga^{\prime}\de^{\prime}}&=&
\frac{1}{4} \left(1-\frac{2}{N}+\frac{1}{N^2} \right)
\Psi_{\al\ba\ga\de}
\textrm{~~(diquark-antidiquark channel)}
\nonumber\\
T^a_{\al\al^{\prime}}T^a_{\de^{\prime}\de}
T^b_{\ba\ba^{\prime}}T^b_{\ga^{\prime}\ga}
\Psi_{\al^{\prime}\ba^{\prime}\ga^{\prime}\de^{\prime}}&=&
\frac{1}{4}\left(N^2+N-2-\frac{2}{N}+\frac{2}{N^2}\right)
\Psi_{\al\ba\ga\de} 
\textrm{~~(meson-meson channel)}
\eea
Inspecting the equation \eq{eq:bs1}, we notice that the energy and
three-momentum integrations separate since the temporal gluon propagator is
energy independent, whereas the heavy quark propagator is independent of the
spatial momentum. Fourier transforming the spatial part of the vertex
\be
\G_{s}(\vec{p_1},\vec{p_2},\vec{p_3},\vec{p_4})=
\int \dk{\vec x_1}\dk{\vec x_2} \dk{\vec x_3} \dk{\vec x_4}
\exp(-i{\scriptstyle\sum\limits_{i=1}^4}\vec p_i\cdot\vec x_i)
\G_{s}(\vec{x}_1,\vec{x}_2,\vec{x}_3,\vec{x}_4),
\ee
we find that the convolution product with the temporal gluon propagator is
given by
\be
\int\dk{\vec k} W_{\si\si}(\vec k)
\G_{s}( \vec p_1+\vec k,\vec p_2-\vec k, \vec  p_3,\vec  p_4)=
\int \dk{\vec x_1}\dk{\vec x_2} \dk{\vec x_3} \dk{\vec x_4}
\exp(-i{\scriptstyle\sum\limits_{i=1}^4}\vec p_i\cdot\vec x_i)
W_{\si\si}(\vec x_2-\vec x_1)
\G_{s}(\vec{x}_1,\vec{x}_2,\vec{x}_3,\vec{x}_4)
\ee
and hence the spatial component of the vertex completely drops from
the calculation.  Motivated by the symmetry of the system, we further
restrict to the case of equal quark separations $|\vec r|= |\vec
x_i-\vec x_j|$ $ (i,j=\overline{1,4}; i>j)$. The original equation can
be recast into an equation for the temporal component $\G_t$:
\bea
\lefteqn{\G_{t}(p_1^0,p_2^0,p_3^0,p_4^0) }\nonumber\\
&=&g^2 W_{\si\si}(\vec r)\int\dk{k}\left\{
\frac{1}{3}\left[W_{\bar qq}(p_1^0+k_0) W_{\bar qq}(p_2^0-k_0)
\,\G_{t}(p_1^0+k_0,p_2^0-k_0,p_3^0,p_4^0) +(3,4)\right]
\right. \nonumber\\
&-& \frac{5}{6}\left[W_{\bar qq}^T(p_3^0-k_0) W_{\bar qq}(p_2^0+k_0)
\,\G_{t}(p_1^0,p_2^0+k_0,p_3^0-k_0,p_4^0)+(1,4)+(2,4)+(1,3)\right] 
\bigg\}\nonumber\\
&-&\!\!(g^2 W_{\si\si}(\vec r))^2\!\!\!\int \!\!\dk{k}\dk{q}
\left\{
\frac{1}{9}W_{\bar qq}(p_1^0+k_0) W_{\bar qq}(p_2^0-k_0)
W_{\bar qq}^T(p_3^0+q_0) W_{\bar qq}^T(p_4^0-q_0)
\,\G_{t}(p_1^0+k_0,p_2^0-k_0,p_3^0+q_0,p_4^0-q_0) 
\right. \nonumber\\
&+&
\left.
\frac{43}{18}\left[W_{\bar qq}(p_1^0+k_0)W_{\bar qq}^T(p_3^0-k_0)
W_{\bar qq}(p_2^0+q_0)W_{\bar qq}^T(p_4^0-q_0)
\,\G_{t}(p_1^0+k_0,p_2^0-q_0,p_3^0-k_0,p_4^0+q_0)
+(1,4;2,3)\right]
\right\}, 
\nonumber\\
\label{eq:bs2}
\eea
where $(i,j)$ represent the terms attached to the corresponding pairs of
(anti)quarks, and can be explicitly read off from Eqs.~(\ref{eq:bs1}, \ref{eq:bspairs}).

Now in order to identify the structure of the solution, it is useful to
rewrite the energy integral as follows:
\bea
&&\int\dk{k_0} W_{\bar qq}(\tilde p_{2}^0-k_0-m) W_{\bar
  qq}(\tilde p_{1}^0+k_0-m)\tilde \G_t(\tilde p_{1}^0+k_0,
 \tilde p_{2}^0-k_0, \tilde p_{3}^0, \tilde p_{4}^0)\nonumber\\
&=&-\frac{2}{\tilde p_{1}^0+ \tilde p_{2}^0-2{\cal I}_r+i\e}
\int\dk{k_0}
\frac{
\G_t( \tilde p_{1}^0+  \tilde p_{2}^0+k_0,-k_0, \tilde p_{3}^0,
\tilde p_{4}^0)}
{\tilde p_{1}^0+  \tilde p_{2}^0+k_0-{\cal I}_r+i\e},
\label{eq:energy}
\eea
where we have introduced the shifted momenta $\tilde p_{1,2}^0=p_{1,2}^0+m$
for notational convenience. The integration over antiquark propagators leads
to an identical formula, except that the mass term has the opposite sign (in
this case, $\tilde p_{3,4}^{0}=p_{3,4}^{0}-m$), whereas in the integral over a
quark and an antiquark, the mass term completely vanishes. The double
integrals can straightforwardly be rewritten in a similar form. Without loss
of generality, we can further restrict to equal energies, i.e., $\tilde
p_i^0=P_0/4$. Inspired by the energy integral \eq{eq:energy}, and noticing the
similarities with our previous three-body calculation \cite{Popovici:2010ph},
we make the following Ansatz for the tetraquark vertex
\be
\G_t(\tilde p_{1}^0,\tilde p_{2}^0, \tilde p_{3}^0,\tilde p_{4}^0)=
\sum\limits_{\substack {\mathclap{i,j=\overline{1,4}}\\\mathclap{i<j}}}
\;\;\frac{1}{\tilde p_i^0+\tilde p_j^0-2{\cal I}_r-A(P_0,{\cal I}_r)+i\e}
\ee
where $A(P_0,{\cal I}_r)$ is a function that needs to be determined. For equal
energies, the Ansatz takes the simpler form:
\be
\G_t(\tilde p_{1}^0,\tilde p_{2}^0,\tilde p_{3}^0,\tilde
p_{4}^0)\big|_{\tilde p_i^0=\frac{P_0}{4}}=\frac{12}{P_0-4{\cal
    I}_r-2A(P_0,{\cal I}_r)+i\e}.
\label{eq:ansatz}
\ee
Plugging this back into \eq{eq:bs2} and using the result \eq{eq:tempint}, we
are left with an algebraic equation for the function $A(P_0,{\cal I}_r)$.  As
has been emphasized in \cite{Popovici:2010mb}, there are only two
possibilities for the bound state energy once all regulators are removed:
either it is finite and linear rising with distance (i.e. a confined state),
or it is infinite and therefore unphysical. Since we are searching for a
confining solution for our tetraquark, the following condition has to be
satisfied:
\be
P_0-4{\cal I}_r=2 C_F i g^2 W_{\si\si}(\vec r)\,.
\label{eq:boen}
\ee
This condition essentially requires that the Fourier transform integral is
convergent, such that the bound state energy remains finite:
\be
\int \dk{\w}\frac{1}{\vec\w^4}(1-e^{i\vec\w\vec r})=\frac{|\vec r|}{8\pi}.
\ee
Replacing ${\cal I}_r$ and $ W_{\si\si}$ by their expressions \eq{eq:inco} and
\eq{eq:Wsisi}, respectively, and the gluon dressing function with
$D_{\si\si}=X/\vec\w^2$, we can rewrite \eq{eq:boen} as
\be
P_0 \equiv \sigma_{4q}|\vec r| = \frac{g^2 C_F X}{4\pi} |\vec r|.
\label{eq:boen1}
\ee
Inserting the Ansatz~(\ref{eq:ansatz}) into \eq{eq:bs2} we find, after a
laborious but fairly straightforward calculation,
\be
A(P_0,{\cal I}_r)=\frac{5}{4 C_F}(P_0-4 {\cal I}_r),
\ee
which gives for the temporal component of the tetraquark vertex
function:
\be
\G_t(\tilde p_{1}^0,\tilde p_{2}^0,\tilde p_{3}^0,\tilde
p_{4}^0)= \sum\limits_{\substack {\mathclap{i,j=\overline{1,4}}\\\mathclap{i<j}}}
\;\;\frac{1}{\tilde p_i^0+\tilde p_j^0-\frac{15}{16}P_0
+\frac{7}{4}{\cal I}_r+i\e}.
\ee

These results are inline with our previous findings for $\bar qq$ and $3q$
systems. From \eq{eq:boen1} we find that the 'string tension' $\sigma_{4q}$
corresponding to a tetraquark state, i.e. the coefficient of the four-body
linear confining term, is two times bigger than the one of $\bar qq$ system
calculated in Ref.~\cite{Popovici:2010mb}:
\be
\sigma_{4q}=\frac{g^2C_F X}{4\pi}=2\sigma_{\bar qq}.
\ee 
For comparison, the string tension for three quark states has the value
$\sigma_{3q}=\frac{3}{2} \sigma_{\bar qq}$. Just like in the case of meson
and baryon states, our results show that in Coulomb gauge and at leading order
in the mass expansion there is a direct connection between the string tension
and the nonperturbative \YM sector of the theory, at least under the
truncation considered here. Notice also that the total mass has disappeared
(similar to mesons), since the two quarks and two antiquarks move with
opposite (and equal) velocities such that the center of mass is stationary. In
fact this is related to our original specification for the Feynman
prescription -- recall that we have assigned the reversed sign for antiquarks,
which corresponds to a particle that moves with opposite velocity.  For
comparison, in the case of baryons, where the three quarks move in the same
direction with equal velocities, the total bound state energy contains three
times the quark mass.

\section{Summary and conclusions}

In this paper we have derived the four-quark confinement potential in the
heavy quark limit of Coulomb gauge QCD. To this end, we have solved the
Yakubovsky equation for tetraquark states in a symmetric configuration and for
equal quark energies.  We have expanded the QCD action by using a method
adapted from HQET, and restricted to the leading order. Further, we have
truncated the system such that only nonperturbative propagators of the \YM
sector are included, and all pure \YM vertices and higher order functions are
neglected.

As in the meson and baryon cases, a direct connection between the physical
string tension and the \YM sector of Coulomb gauge QCD (the temporal gluon
propagator) has been established.  A bound state energy that raises linearly
with the distance has been derived, and the coefficient of the linearly rising
term is found to be two times that of a meson system.  Since only symmetric
configurations have been considered, no statement can be made regarding the
shape of the string that confines the quarks. However, the restriction to
equal energies does not alter the validity of our statements -- clearly the
confining potential should hold for any configuration, including that of equal
energies.

A possible extension of this work is to include the next order in the mass
expansion, and analyze the contribution of the spatial gluon propagator which
so far has been neglected. A different line of research is the inclusion of
vertex corrections -- this should trigger the phenomenon of charge screening
which is expected to alter the value of the string tension. Finally, our
result serves as a basis for phenomenological descriptions of heavy
tetraquarks in terms of potentials.

\begin{acknowledgments}
We are grateful to Peter Watson for useful discussions and a critical reading
of the manuscript. This work was supported by BMBF under contract 06GI7121 and
by the Helmholtz International Center for FAIR within the LOEWE program of the
State of Hesse.
\end{acknowledgments}

\bibliography{$HOME/bibliography/biblio}

\end{document}